\documentclass[aps,pra,preprint,groupedaddress,showpacs]{revtex4}
\newcommand{\bra}[1]{\langle {#1} |}
\newcommand{\ket}[1]{| {#1} \rangle}

\newcommand{\Ham}{{\cal H}}
\newcommand{\vecr}{{\mathbf r}}
\newcommand{\vecR}{{\mathbf R}}
\usepackage{amsmath,graphicx,bm}
\begin{document}

\title{Application of density functional theory to line broadening: Cs
atoms in liquid helium}

\author{Takashi~Nakatsukasa}
\email[Electronic address: ]{takashi@nucl.phys.tohoku.ac.jp}
\affiliation{Physics Department, Tohoku University, Sendai 980-8578, Japan}

\author{Kazuhiro~Yabana}
\email[Electronic address: ]{yabana@nucl.ph.tsukuba.ac.jp}
\affiliation{Institute of Physics, University of Tsukuba,
Tennodai 1-1-1, Tsukuba 305-8571, Japan}

\author{George~F.~Bertsch}
\email[Electronic address: ]{bertsch@phys.washington.edu}
\affiliation{Physics Department and Institute for Nuclear Theory,
University of Washington, Seattle, WA 98195, USA}

\date{\today}

\begin{abstract}
We test the applicability of density functional theory (DFT) to
spectral perturbations taking an
example of a Cs atom surrounded by superfluid helium.
The atomic DFT of helium is used to obtain the distribution
of helium atoms around the impurity atom, and the electronic DFT is
applied to the excitations of the atom,
averaging over the ensemble of helium configurations.
The shift and broadening of the
$D_1$ and $D_2$ absorption lines are quite well reproduced by theory,
suggesting that the DFT may be useful for describing
spectral perturbations in more complex environments.
\end{abstract}

\pacs{32.70.Jz, 67.40.Yv, 71.15.Mb}

\maketitle

\section{Introduction}

The time-dependent density-functional theory has proven to be
a powerful tool in the description of optical absorption for 
molecules and clusters in free space (For a review, see
Ref. \cite{Ca96}.  More recent citations can be found in
Refs. \cite{YB99,NY01}.).
We would like to know whether the theory can be extended
to calculate the line shifts and broadening when the absorber is
embedded in  a medium.
Most applications have considered the electronic excitation from
a single frozen nuclear configuration representing the ground
state of an isolated molecular or an ideal crystal, in which case
the predicted absorption lines below the photoionization threshold
are sharp. Recently it was shown that the TDDFT 
also works well in describing
the broadening of the transitions and the strength
of symmetry-forbidden transitions due to zero-point vibrational
motion, taking the example of the benzene spectrum \cite{Be01}.  In the
present work, we 
will calculate the effects of external perturbations on the optical
absorption.
We choose as a simple test case a Cs atom immersed in liquid helium
at low temperature, because of the simplicity both of the
electronic structure and of the external perturbation.  In the long term we
are interested in extending this kind of analysis 
more complex systems which would
require the full power
of the time-dependent density functional theory.

Aside from our motivation from the perspective of applications
of DFT, spectroscopic measurements of impurity atoms and 
molecules in superfluid helium have been attracting considerable 
interest in recent years \cite{TV98}.
The repulsive force between an impurity and 
helium atoms induces a ``bubble'' around the impurity. This leads to
a weak perturbation of helium atoms on the spectra of impurities.
The line shifts and spectral shapes induced by the helium perturbation 
provide information on the properties of the bubble in the quantum liquid
as well as the 
excited states of the impurity. Since the perturbation is weak, 
this method also provides a unique opportunity to spectroscopic
measurements 
of atomic clusters at low temperature \cite{PHNT95,HEC96,BCFQT96}.

The perturbations of Cs lines have been studied experimentally
as a function of helium density \cite{KFTY95} and we shall calculate this
system. There are two $s$-to-$p$
transitions, the
$D_1$ ($s_{1/2}\rightarrow p_{1/2}$)
and $D_2$ ($s_{1/2}\rightarrow p_{3/2}$) lines,
which are blue-shifted in helium and acquire widths.
The shifts and widths of the two lines are different, and
the $D_2$ line 
has a skewed shape suggesting a double-peak structure. These features 
were first analyzed with 
a collective vibration model of the helium bubble \cite{KFY96}.
The model reproduced average peak shifts, but gave
line widths less than a half of observed ones.
A more sophisticated analysis has been made treating the  
liquid helium environment by the Path-Integral Monte-Carlo \cite{Oga99}. 
This quantum simulation has succeeded to reproduce observed $D_2$
line profiles of the Cs spectra.  
However, this computation method is very costly in computer resources.
We are thus 
motivated to develop an approach to treat helium perturbation that is both
simple and yet has quantitative accuracy.
We will show that a density functional theory (DFT) together with a 
statistical description of helium configurations 
meets our purpose. The helium density 
distribution around an embedded atom is calculated
with the DFT, and the 
helium configurations are generated by a random sampling using the 
density distribution as the sampling weight function. 

Besides reporting the calculations on the absorption spectra of
Cs in helium, we offer some simple qualitative 
interpretation of how the qualitative features of the spectrum
reflect the properties of the helium bubble around the atom.

\section{Formalism}

\subsection{Description of liquid helium
            around an impurity atom}

Among number of density functional methods for liquid helium,
we adopt the Orsay-Paris functional of Ref. \cite{Dup90}.
Although the Orsay-Paris functional is known to have some problems with
dynamic properties of liquid helium \cite{PT94},
it has a  correct long-range behavior and reasonable
short-range characteristics.
Since we are interested in a density profile of liquid helium,
it should be adequate for our purposes.

The energy in the DFT is assumed to 
have the form
\begin{equation}
E=\int d\vecr \Ham_0(\vecr) ,
\end{equation}
where
\begin{equation}
\label{H_0}
\Ham_0(\vecr) = \frac{1}{2m}\left| \nabla\sqrt{\rho(\vecr)} \right|^2
 + \frac{1}{2}\int d\vecr' \rho(\vecr)\rho(\vecr') V_{\rm LJ}(|\vecr-\vecr'|)
 +\frac{c}{2}\rho(\vecr)\left(\bar\rho_\vecr\right)^{1+\gamma} .
\end{equation}
Here, $m$ is the mass of a helium atom and $\bar\rho_\vecr$ is a coarse-grained density defined by
\begin{equation}
\label{rho_bar}
\bar\rho_\vecr=\frac{3}{4\pi h^3} \int_{r<h}d\vecr \rho(\vecr) .
\end{equation}
The $V_{\rm LJ}$ is a standard Lennard-Jones potential describing the
He-He interaction screened at distances shorter than the distance $h$,
\begin{equation}
\label{V_LJ}
  V_{\rm LJ}(|\vecr-\vecr'|) =\begin{cases}
  4\epsilon \left[ \left( \frac{\alpha}{|\vecr-\vecr'|}\right)^{12}
                 -\left( \frac{\alpha}{|\vecr-\vecr'|}\right)^6 \right] ,
       & \mbox{for } |\vecr-\vecr'| \geq h ,\cr
  V_{\rm LJ}(h) \left(\frac{|\vecr-\vecr'|}{h}\right)^4 ,
       & \mbox{for } |\vecr-\vecr'| < h .
\end{cases}
\end{equation}
The values of the parameters in Eqs. (\ref{H_0}), (\ref{rho_bar}), 
and (\ref{V_LJ}) are
$c=1.04554\times 10^7$ K\AA$^{3(1+\gamma)}$,
$\gamma=2.8$,
$\epsilon=10.22$ K, $\alpha=2.556$ \AA, and $h=2.377$ \AA.
This is the same density functional that was used in Ref. 
\cite{Dal94} to study atomic impurities in liquid helium.
In that work, the effect of the impurity was treated by
including a potential interaction $V_I(r)$ between it and the helium
atoms in the density functional,
\begin{equation}
\Ham(\vecr) = \Ham_0(\vecr) + V_{\rm I}(\vecr)\rho(\vecr) ,
\end{equation}
The $V_I(r)$ has important contributions from
the repulsive  
between electrons and the helium atoms, as well
as the Van der Waals-type polarization interaction.  
Since we need to treat interaction of impurity electrons with
the helium atoms explicitly later on when we calculate the
electronic excitation,
we introduce it here as well for calculating the helium
distribution.  We approximate it as a contact interaction, i.e.
of the form
\begin{equation}
\label{V_e_He}
V_{\rm e-He}(\vecr_{\rm e}-\vecr)=V_0 \delta(\vecr_{\rm e}-\vecr) ,
\end{equation}
where $\vecr_{\rm e}$ and $\vecr$ are coordinates of the electron and
helium atom, respectively.
The strength $V_0$ is determined from the electron-helium scattering length
$a$ as
\begin{equation}
V_0=\frac{2\pi a}{m_{\rm e}},
\end{equation}
where $m_e$ is the electron mass.  
Then $V_{\rm I}(\vecr)$ is given by
\begin{equation}
V_{\rm I}(\vecr)
 = V_0 \rho_{\rm e}(\vecr) .
\end{equation}
where $\rho_{\rm e}(\vecr)$ is the electron density of the impurity atom.
We take 
$a=0.69$ \AA, corresponding the observed low-energy electron-helium
cross section $\sigma = 6.0$ \AA$^2$ \cite{Tra96}.
We have also assumed that the ion core is heavy enough to be treated as
a classical particle at the origin.
Since Eq. (\ref{V_e_He}) expresses the interaction between He atoms and
an electron, the same interaction will be used to estimate the energy
shift of valence electrons due to the helium perturbation.
This treatment of interatomic potential ignores long-range attraction
due to the polarization effects. The influence of the polarization effect
will be mentioned later.

\begin{figure}[ht]
\includegraphics[width=0.6\textwidth]{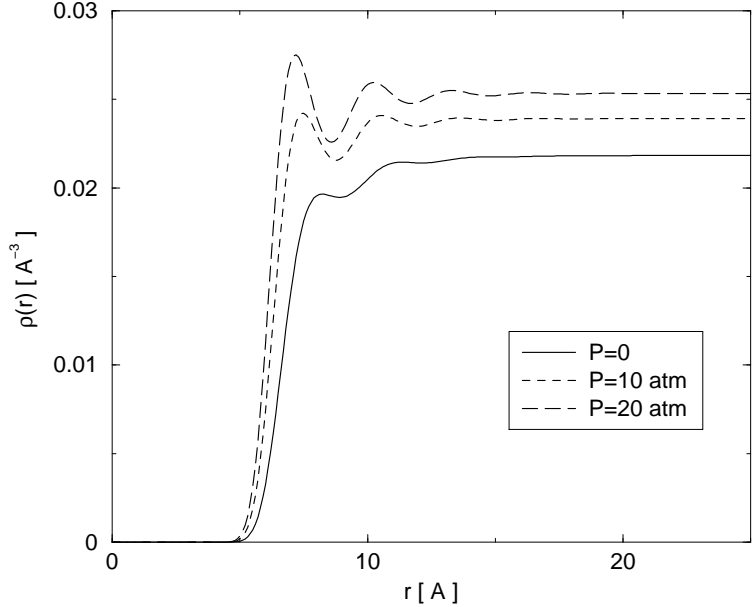}
\caption{\label{fig: Cs_rho}
Helium density profile around Cs atom.
The coordinate $r$ represents the distance from the Cs atom. 
}
\end{figure}

Utilizing the energy functional, $E[\rho]=\int d\vecr \Ham(\vecr)$,
we calculate the density profile of liquid helium, putting the impurity
atom at the origin. Minimizing the grand potential at zero temperature,
$\Omega\equiv E[\rho(\vecr)] - \mu N$, leads to a Hartree-type
equation
\begin{equation}
\label{Hartree}
\left[ -\frac{1}{2m} \nabla^2 + U(\vecr) + V_{\rm I}(\vecr) \right]
  \sqrt{\rho(\vecr)} = \mu \sqrt{\rho(\vecr)} ,
\end{equation}
where
\begin{equation}
\label{chemical_pot}
U(\vecr)\equiv \int d\vecr' \rho(\vecr') V_{\rm LJ}(|\vecr-\vecr'|)
  +\frac{c}{2} \left(\bar\rho_\vecr \right)^{\gamma+1}
  +\frac{c}{2}(1+\gamma)\frac{3}{4\pi h^3}
   \int_{|\vecr-\vecr'|<h} d\vecr'
       \rho(\vecr') \left( \bar\rho_{\vecr'} \right)^\gamma .
\end{equation}
The equation is solved with the boundary condition that the density
go to the bulk density $\rho_0$ at large $\vecr$, which can be satisfied
setting the chemical potential to
\begin{equation}
\mu = b\rho_0+\left(1+\frac{\gamma}{2}\right)c \rho_0^{\gamma+1} ,
\end{equation}
where
$
b=\int d\vecr V_{\rm LJ}(\vecr) = -8.8881\times 10^2 \mbox{ K\AA}^3$.
The bulk density is related to the pressure
$P$ by 
\begin{equation}
\label{pressure}
P=-\frac{\partial E}{\partial V} = \frac{1}{2}b\rho_0^2
  + \frac{\gamma+1}{2}c\rho_0^{\gamma+2} .
\end{equation}
Carrying out the solution of Eq. (\ref{Hartree}) we find the
density profile shown in Fig. \ref{fig: Cs_rho}.
The three curves give $\rho(r)$ at equilibrium densities $\rho_0$ of
0.0218, 0.0239, and 0.0253 \AA$^{-3}$, corresponding to 
$P=0$, 10, and 20 atm, respectively.  One can see a sharp rise in
the density at $r\approx 6$ \AA.  This corresponds to the bubble
radius.
An oscillatory structure appears on the density profile,
especially under high pressure.
This feature is different from that of a bubble model adopted
in Refs.~\cite{KFTY95,KFY96}.
The maximum value of the density for $P=20$ atm is
about $0.0275$ \AA$^{-3}$ at $r=7.2$ \AA.

We use the $\rho(r)$ computed above to generate configurations of 
helium atoms as follows.
Take a large volume surrounding the alkali atom and 
denote it as $V$. This volume includes $N$ helium atoms in average,
where $N$ is given by $\int_V d\vecr \rho(\vecr) = N$. 
We randomly sample $N$ helium positions in $V$ according to the density 
distribution $\rho(\vecr)$. This sampling procedure gives probability 
distribution without correlation among helium atoms. Denoting
$f(\vecr) = \rho(\vecr)/N$, the probability distribution of $N$ atoms
is given by
\begin{equation}
w_{\rm nc}({\bm{\tau}}) = \prod_{i=1}^N f(\vecr_i) ,
\end{equation}
where $\bm{\tau}$ stands for $(\vecr_1,\cdots,\vecr_N)$.
We will also study the effect of helium-helium correlations by considering
a probability distribution of the form
\begin{equation}
\label{correlated_prob}
w_{\rm c}(\bm{\tau}) = \prod_{i=1}^N g(\vecr_i)
\prod_{i<j}^N \theta(r_{ij}-d).
\end{equation}
Here $d$ is the range of a short-range correlation.
The distribution function $g(\vecr)$ is determined
by the condition that the distribution of Eq.~(\ref{correlated_prob})
gives a helium density $\rho(\vecr)$,
\begin{equation}
\label{def_fc}
\rho(\vecr)=N\int_V d\vecr_2 \cdots d\vecr_N
w_{\rm c}(\vecr,\vecr_2,\cdots,\vecr_N) .
\end{equation}
In practice, we employ an iterative procedure to find $g(\vecr)$ from this
condition.
\subsection{Helium perturbation on the atomic spectra}

In the previous section, we described a density functional theory
for calculating the density profile of liquid helium $\rho(\vecr)$,
and taking account of the effect of an impurity atom at the origin.
In this section, we discuss the calculation of the atomic spectrum,
including the effects of the helium atoms.

\begin{figure}[ht]
\includegraphics[width=0.5\textwidth]{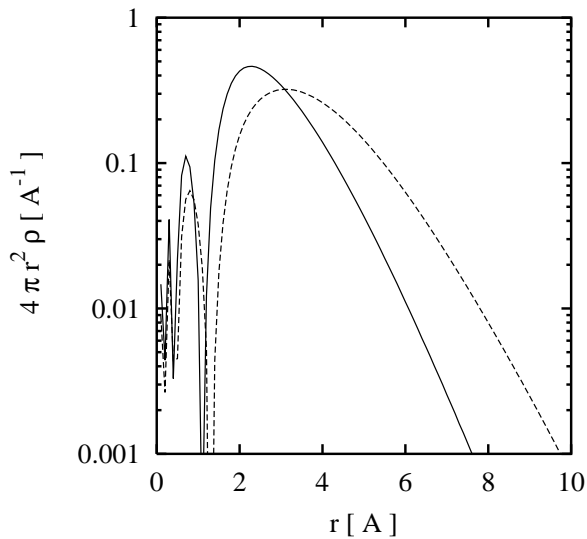}
\caption{\label{fig: elect}
Electron density distribution of the $s_{1/2}$ (solid) and $p_{1/2}$
orbitals of Cs.  See text for details of the calculation.
}
\end{figure}

We begin with the theory of the isolated atom.  Orbital wave functions
are calculated using density functional theory
with Dirac wave functions and kinetic energy operator.  
We need accurate wave functions at large distances from the atom,
which cannot be achieved with the traditional LDA functional due
to the incorrect orbital eigenvalues and the incorrect asymptotic
behavior of the potential.  As is well known, these problems
are diminished with the GGA density functionals.  We employ
the GGA functional of Ref. \cite{LB94}, which has a gradient
correction to produce correct asymptotic behavior of the potential.
The gradient correction includes an adjustable 
parameter $\beta$; we utilize this freedom to make the orbital
energy coincide with the measured one. For the $s_{1/2}$ orbital, the
orbital energy is set equal to the ionization potential of Cs atom,
3.89 eV.
For the $p_{1/2}$ and $p_{3/2}$ orbitals, the orbital energies are set
equal to the ionization potential minus the excitation energies (about
1.43 eV).
The quality of the wave function may be tested by examining the
transition oscillator strength. For $D_1$ and $D_2$ transitions,
the calculated oscillator strength assuming a pure single-electron
transition is 1.034, in good agreement with measured value, 1.058.
The calculated electron density distributions are shown in
Fig.~\ref{fig: elect}.
In principle, there will be contributions to the transition from
core electrons as well that can be taken into account with the
TDDFT.  Applying the TDDFT to the present case, 
the core contributions reduce the oscillator strength by some tens
of percent, but do not significantly affect the asymptotic wave function 
of the valence electron.  We therefore use the simpler single-electron
wave functions below rather than wave functions from the TDDFT.

For a given configuration of helium atoms,
$\bm{\tau}=(\vecr_1,\cdots,\vecr_N)$,
we evaluate shifts of orbital energies with first order perturbation 
theory. We use the same helium configuration for the ground state $s_{1/2}$
and excited states $p_{1/2}$ and $p_{3/2}$, following the Frank-Condon
principle. For $s_{1/2}$ and $p_{1/2}$ states, the energy shifts of the
valence electron is calculated as
\begin{equation}
\Delta E^{(k)}(\bm{\tau}) = \langle \psi^{(k)} \vert
\sum_i V_{\rm e-He}(i) \vert \psi^{(k)} \rangle
=\sum_i V_0 |\psi^{(k)}(\vecr_i)|^2
\end{equation}
where $k$ stands for orbital quantum numbers $(\ell j)$ and either
$m$ state may be taken.  For $p_{3/2}$ states, the matrix elements 
depend on $m$ and we have to diagonalize a $4 \times 4$ matrix to get
the energy shifts.  We then obtain two eigenenergies, each of which is doubly
degenerate.

Each helium configuration produces an energy shift and possible splitting
but the transitions remain sharp.  The line broadening
comes from the ensemble average over helium configurations.
The line shape of the spectra for $D_1$ ($s_{1/2} \rightarrow p_{1/2}$)
transition is given by
\begin{equation}
S_{D_1}(E) = \int_V d\bm{\tau}
w(\bm{\tau}) 
\delta \left( E - \left( \Delta E^{(p_{1/2})}(\bm{\tau}) 
- \Delta E^{(s_{1/2})}(\bm{\tau}) \right) \right) ,
\label{defspectrum_1}
\end{equation}
where $E$ is a shift from the energy position of the free atom.
For the $D_2$ ($s_{1/2} \rightarrow p_{3/2}$) transition,
we need to add the two eigenmodes,
\begin{eqnarray}
S_{D_2}(E) = \int_V d\bm{\tau}
w(\bm{\tau}) \left\{
\delta \left( E - \left( \Delta E_1^{(p_{3/2})}(\bm{\tau}) 
- \Delta E^{(s_{1/2})}(\bm{\tau}) \right) \right) \right.\nonumber \\
\left. +\delta \left( E - \left( \Delta E_2^{(p_{3/2})}(\bm{\tau})
- \Delta E^{(s_{1/2})}(\bm{\tau}) \right) \right) \right\} .
\label{defspectrum_2}
\end{eqnarray}

\subsection{Origin of peak shifts, broadening, and splitting}

In this section, we discuss qualitatively argument how the blue shifts,
the broadening of the lines, and the 
splitting of $D_2$ transitions occur.
First consider a valence electron of the alkali atom interacting
with a single helium atom at a position $\vecR$ with respect to the
impurity atom.
The energy shift of the valence electron is
determined by the electron wave function at the position of helium atom,
$\psi(\vecr_{\rm e}=\vecR)$.

For $s_{1/2}$ and $p_{1/2}$ states, the energy shift $\Delta\epsilon$ is
calculated as
\begin{equation}
\label{shift_1/2}
\Delta\epsilon^{(k)} 
 = V_0 \rho_{\rm e}^{(k)}(R) ,
\end{equation}
as the first-order perturbation by $V_{\rm e-He}$.
Since $V_0$ is positive, the energy shift, Eq. (\ref{shift_1/2}), is
also positive.  As may be seen from Fig. \ref{fig:  elect}, 
the wave function of the $p_{1/2}$ state is considerably
larger than the $s_{1/2}$ state outside the bubble.
Thus, we expect a blue shift of $D_1$ excitation spectra,
$\Delta\epsilon^{(p_{1/2})}-\Delta\epsilon^{(s_{1/2})}>0$.

For $p_{3/2}$ states, the situation is slightly more complicated,
because, in general, there are off-diagonal matrix elements among
degenerate states with different $m$.
However, all the off-diagonal elements vanish if we assume that
the helium atom lies on the $Z$-axis, $X=Y=0$ and $Z=R$.
We lose no generality in the case of a single helium atom.
The helium atom at $Z=R$ produces an energy shift $\Delta\epsilon$ as
\begin{equation}
\label{shift_3/2}
\Delta\epsilon^{(p_{3/2})}_m
 = \bra{\psi^{(p_{3/2})}_m} V_{\rm e-He} \ket{\psi^{(p_{3/2})}_{m}}
 = \begin{cases}
    2V_0\bar\rho_e^{(p_{3/2})} ,
       & \mbox{ for }|m|=1/2, \\
    0, & \mbox{ for }|m|=3/2.
 \end{cases}
\end{equation}
where $\bar\rho^{(k)}$ is the angle-average electron density.
There is no shift for $|m|=3/2$ states in the first-order
perturbation by $V_{\rm e-He}$.
As a result, the $D_2$ transition splits into two peaks;
one has a blue shift,
$\Delta\epsilon^{(p_{3/2})}_{|m|=1/2}-\Delta\epsilon^{(s_{1/2})}>0$,
and the other has a small red shift, $-\Delta\epsilon^{(s_{1/2})}<0$.
In fact we can neglect  $-\Delta\epsilon^{(s_{1/2})}$ in comparison 
to the other shifts due to the large distance of the perturbing helium
atom.  Also one can neglect the small energy difference between the
two spin orbit partners.  Then the
energy shift of $|m|=1/2$ states is twice of that of
$p_{1/2}$ states.  Note that the average shift of the $D_2$ 
components is the same as the $D_1$ line.
In summary, a helium atom at a distance $R$ shifts
the $D_1$ transition by $\Delta\equiv V_0 \rho_e^{(p_{1/2})}(R)$.
The energy of $D_2$ transition splits into two: one has no shift and
the other is shifted by $2\Delta$.  This explains qualitatively the 
overall blue shift of both lines and the skewed profile of the
$D_2$ line.  

However, while the observed $D_2$ line shape can be analyzed as a sum of two
components, both components are blue shifted roughly the same amount
as the $D_1$ line.  This shows the shift is due to the simultaneous
interaction with several helium atoms.
For example, if two helium atoms are at $(R,0,0)$ and at $(0,0,R)$,
again, a $D_2$ line splits into two.
However, in this case, both have blue shifts ($\Delta$ and $3\Delta$).
If three helium atoms are at $(R,0,0)$, $(0,R,0)$, and $(0,0,R)$,
then, $D_2$ line shows a blue shift of $3\Delta$ but no splitting.
Roughly speaking,
the magnitude of line shift is determined by number of helium atoms
contributing to the perturbation
and the splitting is determined by anisotropy of the helium configuration
which increases as the square root of the number of atoms, assuming
that there are no correlations between atoms.
The average energy shift remains the same for the $D_1$ and $D_2$
lines, irrespective of the number of helium atoms causing the
perturbation.


\section{Numerical results}

\begin{figure}[ht]
\includegraphics[width=0.8\textwidth]{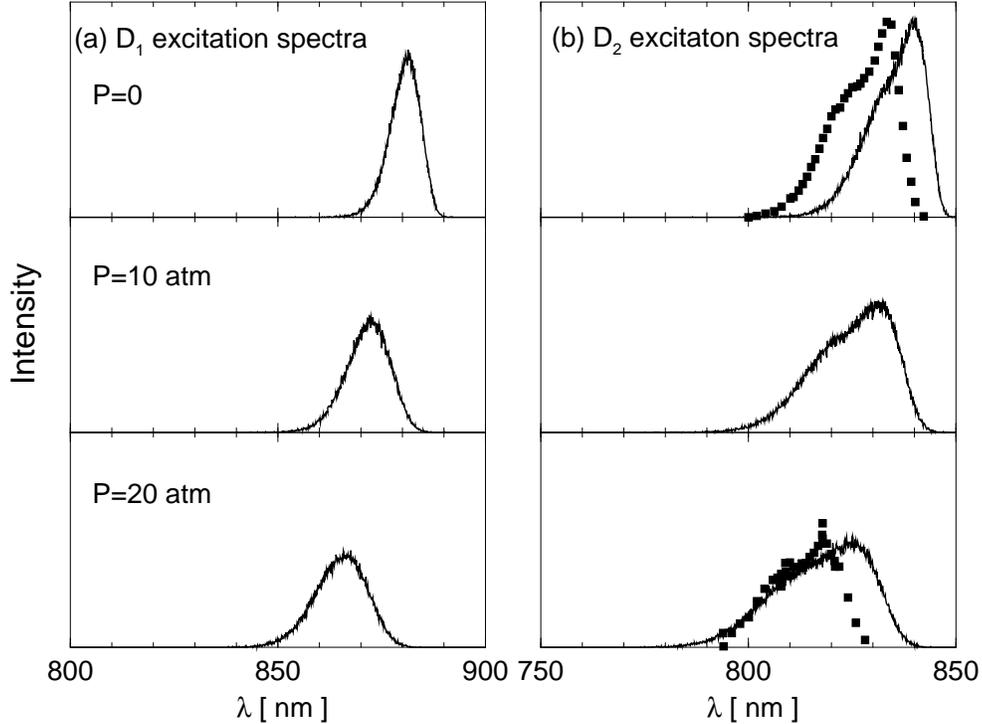}
\caption{\label{fig: No_correlation}
(a) Cs $D_1$ excitation spectra at different helium pressure;
$P=0$, 10, and 20 atm
(b) The same as (a) but for $D_2$ excitation spectra.  Experimental
data from Ref. \cite{KFTY95} are plotted as filled squares.
}
\end{figure}

To calculate line shapes of the Cs $D$ transition in a helium,
we evaluated Eq. (2.18) and (2.19) by sampling 100 000 helium 
configurations, generated according to
the density profiles in Fig.~\ref{fig: Cs_rho}.
The calculated energy shifts are added to
the observed $D$ lines of free Cs atom
($\lambda=894.9$ nm for $D_1$ and 852.7 nm for $D_2$).
Then, the intensity is estimated by counting number of events
in bins of wavelength $\Delta\lambda=0.1$ nm.
The obtained intensity spectra are shown in Fig.~\ref{fig: No_correlation}.
The $D_1$ line can be well approximated by a single Gaussian,
although there is a slightly larger tail at high-energy
(low-wavelength) side.
On the other hand, the $D_2$ line has a double-peaked structure.
This feature agrees with the experimental observation \cite{KFTY95,KFY96}.

\begin{figure}[ht]
\includegraphics[width=0.5\textwidth]{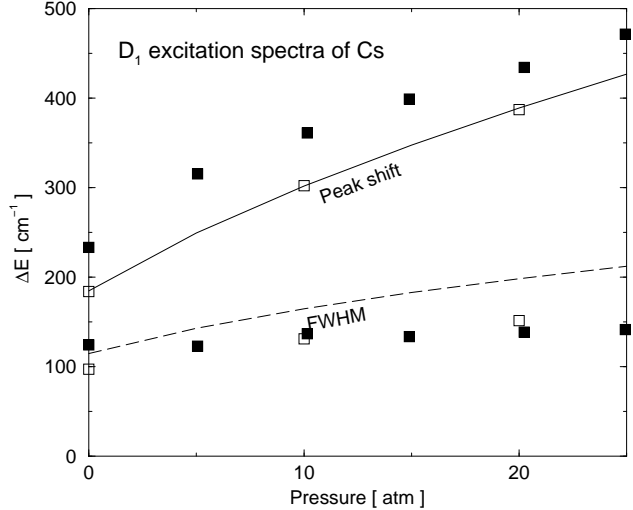}
\caption{\label{fig: D1_P}
Pressure shift (solid line) and broadening (dashed line)
of the $D_1$ excitation
spectra of Cs as a function of helium pressure.
Open squares represent broadening calculated with He-He correlations
at $P=0$, 10, and 20 atm.  The experimental data are plotted as
filled squares.
}
\end{figure}

Fig.~\ref{fig: D1_P} shows the pressure dependence of the peak
shift and broadening of the $D_1$ excitation lines.
The peak shift reproduces the observed pressure dependence, but
comes out about 20\% lower than measured.  The difference could
be due to an incorrect asymptotic wave function in the Cs atom;
only a 10\% error in the wave function would be required
to explain the difference.  Or the calculated helium bubble might
be too large.  Here, decreasing the size of the bubble by 0.3 \AA\ 
out of 6 \AA\ would be sufficient to produce the measured peak
shift.  We shall return to this later. The line broadening
comes out better than would be
expected, given the quality of agreement for the peak shift.
The agreement here shows that the fluctuations of the helium distribution
are well described by the model adopted.

\begin{figure}[ht]
\includegraphics[width=0.8\textwidth]{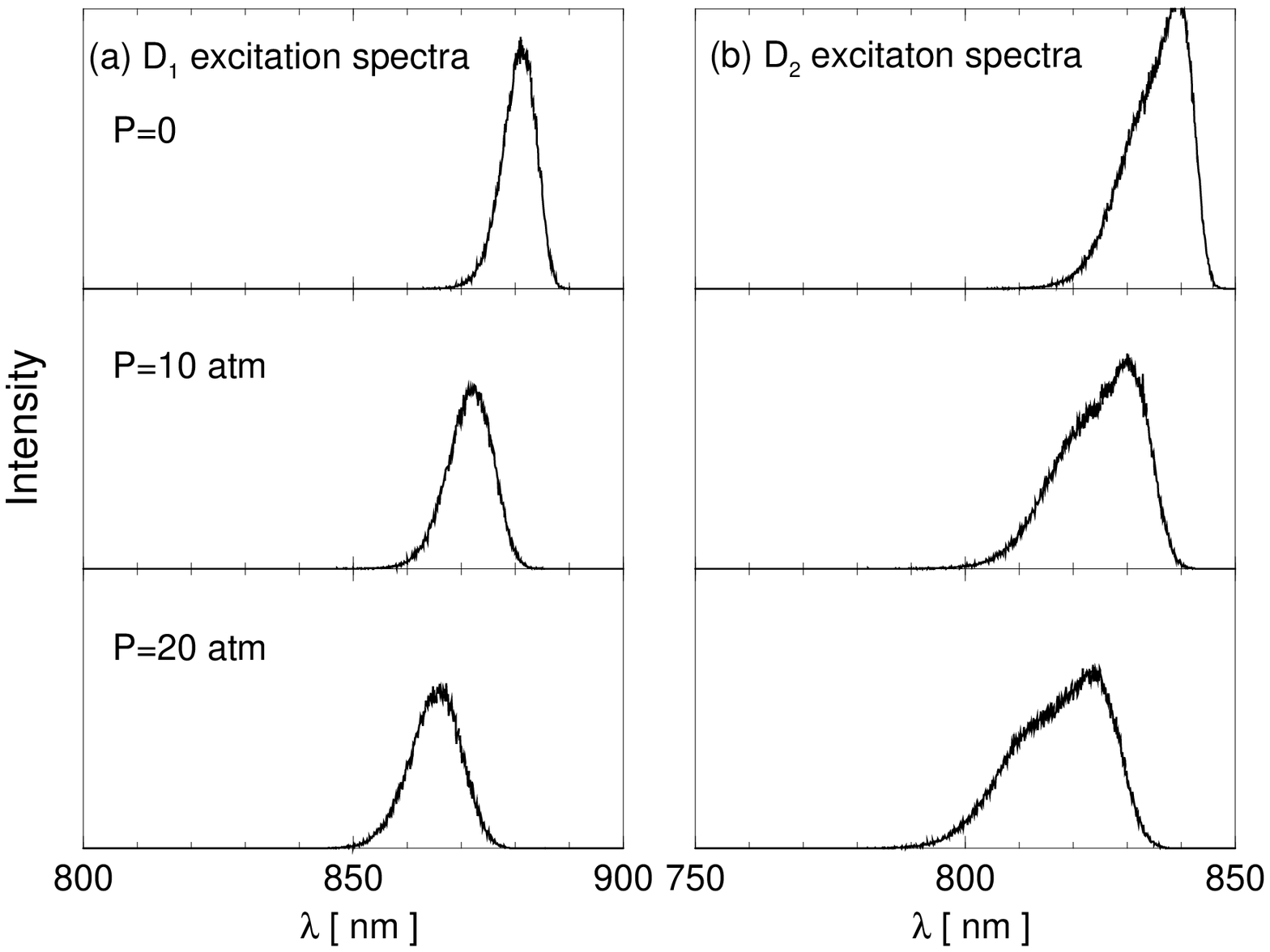}
\caption{\label{fig: correlation}
(a) Cs $D_1$ excitation spectra and (b) $D_2$ excitation spectra
as in Fig. 2, with 
inclusion of He-He correlations.
See text for explanation.
}
\end{figure}

A qualitative measure of the fluctuations can be constructed
by defining an effective number of helium atoms that contribute
to the perturbation.  Calling the shift from an individual helium
atom $\Delta(i)$, the total shift is
\begin{equation}
\Delta E=\sum_{i=1}^N \Delta(i) 
\end{equation}
and the effective number $M_{\rm eff}$ is defined as
\begin{equation}
M_{\rm eff} \equiv \left<
   \frac{\left( \sum_i \Delta(i)\right)^2}{\sum_i \Delta(i)^2}
   \right> ,
\end{equation}
where $\left< \cdots \right>$ indicates the ensemble average.
The calculation leads to $M_{\rm eff}\approx 8$ for $s_{1/2}$ state at
$P=0$, increasing to 11-12 at $P=25$ atm.
The $M_{\rm eff}$ can be compared with number of helium atoms in
the first shell of density profile (Fig.~\ref{fig: Cs_rho}).
The average numbers of helium atoms in a region of $r<8.5\mbox{ \AA}$
are 27 at $P=0$ and 45 at $P=25$ atm.
$M_{\rm eff}$ turns out to be much smaller than the number in
the first shell.
Therefore, we may say that
the perturbation on the valence electron is dominated by
a small number of helium atoms in the inner surface of a bubble.
This fact indicates the importance of treating the
perturbation from individual helium atoms in describing fluctuation
effect.  

We can also understand the order of magnitude of the fluctuation
effects using $M_{\rm eff}$.  Assuming independent
helium atoms, the fluctuations are proportional to $1/\sqrt{M_{\rm eff}}$.
Thus we would expect that widths of the lines and the splitting
of the components of the $D_2$ would be proportional to the average
shift times that quantity.  In fact the $D_1$ width is about 1/2
of its average shift in the zero pressure data, to be compared with
$1/\sqrt{M_{\rm eff}}\approx 1/3$.
Ref. \cite{KFTY95} analyzes the $D_2$ line shape as a sum
two Gaussian peaks.  For the measurements at low pressure, the
splitting of the components relative to the average shift is
just 1/3, agreeing with our very crude argument.

In order to investigate effects of the polarization potential that
we so neglected, we consider also a helium-Cs interaction 
that includes the van der Waals terms  \cite{Pat94}.
Once the density profile $\rho(\vecr)$ is determined,
we use the interaction between a valence electron and helium,
Eq. (\ref{V_e_He}), to calculate the atomic spectra.
We find that potential of Ref. \cite{Pat94} gives a reduced radius 
for the helium bubble at zero pressure, by about 0.3 \AA.
As a result, the blue shift increases by about $35$ cm$^{-1}$ at $P=0$,
which fits the experimental data very well.
The line width is also slightly increased.
This enhancement of blue shifts is decreasing as the pressure is
increasing.
The calculation shows almost no additional shift at P=20 atm.

Next, let us discuss effects of correlations among helium atoms.
The He-He correlation should influence the line width because it
removes some part of fluctuations of the He configuration.
Roughly speaking, the radial fluctuation controls the line width
and the angle fluctuation determines the skewness of $D_2$ line.
To test this, we sample the helium configurations using the
probability distribution of Eq. (\ref{correlated_prob}),
taking $d=2.377$ \AA.
We construct the distribution function $g(\vecr)$ in
Eq. (\ref{correlated_prob}), so as
to reproduce the density profile $\rho(\vecr)$ determined by the DFT,
Eq. (\ref{Hartree}).
The results are displayed in Fig.~\ref{fig: correlation}
and also in Fig.~\ref{fig: D1_P} with open squares.
The correlation effect does not change the average peak positions at all.
However, it reduces the line widths, especially when the liquid helium is
under high pressure.
At $P=20$ atm, the FWHM of $D_1$ line is calculated to be 150 cm$^{-1}$,
which well agrees with the experiment (140 cm$^{-1}$).
The skewness of $D_2$ line also becomes smaller.

\section{Summary}

We have developed a simple model to describe atomic spectra of impurities
embedded in the superfluid helium. Our description employs a density
functional theory for the helium distribution, and treats helium
configurations statistically. The model is applied to the spectra
of Cs atom embedded in superfluid helium. Various features in the spectra,
which include line shifts, broadening, and skewness, are nicely 
reproduced in our calculation without any adjustable parameters. 
Thus we are confident that our model includes basic physical
elements of the helium perturbation correctly.
Since the model is simple enough to apply to more complex systems such
as molecules and clusters, we wish to analyze these systems in future.

\begin{acknowledgments}
One of authors (T.N.) thanks Y.~Matsuo, Y.~Fukuyama,
and R.~Sanetou for discussions.
K.Y. acknowledges financial support by the Ministry of Education,
Culture, Sports, Science and Technology of Japan, Contract
No. 11640372.
G.F.B. acknowledges conversations with P.G. Reinhard and J. Rehr, and
support from the US Department of Energy under
Grant DE-FG06-90ER40561. 
\end{acknowledgments}

\bibliography{draft_6}

\end{document}